\documentstyle [12pt] {article}

\title{
The theory of the centrifugal mechanism of feeding-in in bent crystals
}

\author{Valery Biryukov\thanks{Electronic address:  biryukov@mx.ihep.su  }
\\ Institute for High Energy Physics, \\
 Protvino, 142284 Moscow Region, Russia }

\date{}
\begin{document}
\normalsize
\pagestyle{myheadings}
\markright{\em Physics Letters A {\bf 205}, 343--348 (1995)}

\maketitle

\leftline{Published in: \em Physics Letters A {\bf 205}, 340--343 (1995)}

\begin{abstract}
\normalsize
\baselineskip=15pt
For a particle channeled in the bent crystal planes (axes),
the phenomenon of "bending dechanneling" is well known,
which is a particle transition to random state
due to centrifugal force.
We consider an analytical theory of the reverse phenomenon,
i.e., feeding from a random
to channeled state due to centrifugal force
in a crystal with variable curvature.
\end{abstract}

\normalsize
\baselineskip=15pt

\section{Introduction}

For any trajectory of a particle in a crystal a time-reversed
trajectory is possible.
The starting point of a trajectory
becomes the final one, and vice versa.
This leads  to the idea of {\em reversibility}
of the dechanneling processes \cite{li}.
In the depth of a crystal,  besides the particles
leaving the channeling mode (dechanneling or {\em feeding out}),
there may be particles entering the channeling mode
(respectively, {\em feeding in}, also known as {\em volume capture}).
The mechanisms, responsible for these two opposite processes,
are essentially the same.
Two mechanisms changing the particle state in a crystal,
namely the scattering and the centrifugal effects, are known
(see references in Ref. \cite{ufn}).

In a bent crystal (or a straight crystal with lattice dislocations),
the particle state may be changed by
a change of a curvature (centrifugal effects).
The existence of the trajectories starting in the region of channeled
states and ending outside (or vice versa)
can be shown by solving the equation of particle motion in
a crystal with variable curvature (see, e.g., Ref. \cite{ell84}).
Along these trajectories the particles leave or enter the channeling
mode. The bending dechanneling is a well established phenomenon.
Its theoretical description within the framework of the continuum
model \cite{kudo,ell82} is in excellent agreement with experiment
\cite{forster}.
The same theory inevitably leads to the existence
of the corresponding feed-in process \cite{mannami,bi90}
for the crystal channels with variable curvature
(crystal bending or dislocations).

We shall consider the centrifugal mechanism of feeding-in
by the analytical methods.
The central problem is to find the probability of feeding
in a general case of the charged
particle interaction with the potential
of the crystallographic planes of variable
curvature.
We shall give below two derivations for the probability of feeding:
(1) from the consideration of  particle dynamics,
and (2) from the consideration of reversibility
of trajectories.

\section{Derivation from dynamics}
\label{der1}

Let us consider an efficiency of feeding into
 a planar channel with increasing bending radius R;
the interplanar potential is $U(x)$.
A particle (with momentum $p$ and velocity $v$)
moves as if it were in the effective interplanar potential
\begin{equation}     \label{ueff}
U_{eff}(x) = U(x) + \frac{pv}{R}\cdot x
\end{equation}
with the transverse energy
\begin{equation}     \label{eteff}
E_T = \frac{pv}{2}\theta^2 + U_{eff}(x)
\end{equation}
where $\theta$ is the particle angle relative to the planes
(we denote with $\theta_c$ and $x_c$ the critical angle
and transverse position, respectively).
A change of  curvature
$k=pv/R$ determines, through the relations (\ref{ueff},\ref{eteff})
a corresponding change of the transverse energy
 $E_T$ of a particle  in an effective potential well.
 For rapid variations of $k$,
 when the effective well changes significantly over one period of
 oscillation $\lambda$,
 the term  $E_T$ looses its meaning, and one should
solve the dynamical equation,
for a given function $k(z)$.

In practice, $\lambda$ is quite small ($\sim$30 $\mu$m at 100 GeV),
so the crystal curvature cannot  considerably change
along this length (it is not true for dislocations).
Therefore, the case of a slowly varying curvature
(when the "transverse energy" concept remains useful)
is general enough;
it covers all the bent-crystal experiments done thus far.
In this case, the particle continues to oscillate in the bent channel,
with all the parameters changing slowly.

As a particle moves in the crystal,
the local curvature $k(z)$ varies with the crystal depth $z$.
Instead of the consideration of a variable critical transverse energy
$E_c(z)$, it is more convenient mathematically to keep $E_c$
constant and to introduce an equivalent variation of $E_T$ for
any  particle.
With a small instant variation $\delta k$,
the transverse energy (\ref{eteff}) of the particle
(in a "new" effective well at $z+\delta z$) varies,
with respect to the critical value $E_c$=$U_{eff}(-x_c)$,
by $(x+x_c)\delta k$.
Since  both $\delta k$ and $x$
are dependent on z, the total increase
is $\delta E_T=\int (x+x_c)\delta k$,
where $x$ may depend on the $k(z)$-function
for  rapid variation of $k(z)$.
For slow variations, one may rewrite it as
\begin{equation}	\label{edrift}
\delta E_T=k'\int (x+x_c)dz=k'(\langle x\rangle +x_c)\delta z \; ,
\end{equation}
where we have averaged $x$ over oscillation, and
kept $k'$=$dk/dz$=const along it.

Figure \ref{gap} helps to understand the bending capture.
 The effective interplanar potential is shown at some coordinate $z$
 and at $z+\lambda$.
A random particle may be captured when it is passing
over the potential well with a transverse energy
being in the range from $U_0$ to $U_0+\Delta U$.
The potential well gap
$\Delta U=pvd_p/R$ is due to centrifugal force
($d_p$ is the interplanar spacing).
During oscillation over the well with
$E_T\simeq U_0$, the $E_T$ variation relative to $E_c$ is
\begin{equation}	\label{ecap}
\delta E_T=k'(\langle x\rangle +x_c)\lambda \; .
\end{equation}
Note that $\langle x\rangle$$<$0 for positive $pv/R$.
If $R$ increases, $\delta E_T<0$
and some particles (those with $E_T+\delta E_T<E_c$)
 are captured.

Figure \ref{gap} shows three kinds of particles.
Particles with
         $E_T < U_0+\delta E_T$
are captured; those with
         $E_T > U_0+\Delta U -\delta ' E_T$
 (where  $\delta 'E_T$ is variation
 over $\lambda$/2) are transmitted and interact with
the next wall; other particles are reflected by the channel wall.
For slow variation the number of transmitted particles
may be neglected; then we normalize the captured fraction
to all the particles incident in $\Delta U$ range.
For the flat distribution of transverse energies in the
interval $U_0$ to $U_0+\Delta U$,
the fraction of captured particles is
\begin{equation}     \label{wb}
w_B
=- \frac{\delta E_T}{\Delta U}
=  \frac{R' \lambda}{R} \frac{x_c}{d_p} \cdot
       \left(1+ \frac{\langle x\rangle}{x_c}\right)
\end{equation}
where $R' = dR/dz$ is the curvature radius gradient.
For slightly-bent crystals
the factor in the parentheses is close to unit.
Then one can simplify (\ref{wb}) to
\begin{equation}
w_B   \approx
  \frac{R' \lambda}{2R}
\end{equation}

In the harmonic approximation Eq.\ (\ref{wb}) takes the form:
\begin{equation}     \label{wbh}
w_B
=  \frac{R' \lambda}{R} \frac{x_c}{d_p} \cdot
       \left(1- \frac{R_c}{R}\right)   ,
\end{equation}
$R_c$ being the critical radius.

In the widely-applied 3-point bending scheme, the curvature is a
linear function of $z$: 1/$R(z)$=const$\pm a_0z$.
Interestingly, this leads to a linear dependence of $w_B$
on $R$ (like in the case of 'scattering feeding'
\cite{bir-b73,bir-feed1})
for not too strong bending:
\begin{equation}     \label{wb3p}
w_B =  a_0 \lambda R \frac{x_c}{d_p}
       \left(1+ \frac{\langle x\rangle}{x_c}\right)
       \approx  \frac{a_0 \lambda}{2} R .
\end{equation}
$w_B$ is of order of $2\lambda /L_B$, where $L_B$ is the length of the
bent part of a crystal. The probability grows with energy,
and should be quite sizable, e.g. in conditions of
the H8 experiment \cite{h8-1} at CERN, where (\ref{wb3p})
constitutes several per cent.

\section{Derivation from reversibility}
\label{der2}

The capture probability can be derived  with no use of
approximations, like the existence of
$E_T$, or a slow
varying curvature (although these are well justified in all
practical cases).

Let us consider a beam with a uniform angular distribution,
1/2$\Phi$, incident on a crystal with a curvature
gradually {\em increasing}
along the beam path from 0 to 1/$R_c$.
The fraction of beam channeled will decrease
from the starting value given by
\begin{equation}	\label{b0}
f = \frac{2x_c}{d_p}
    \frac{\pi}{4}
    \frac{\theta_{c}}{\Phi} \ \  ,
\end{equation}
with the crystal depth, according to:
\begin{equation}	\label{bena}
f(z)
 = \frac{2x_c}{d_p}
    \frac{\pi}{4}
    \frac{\theta_{c}}{\Phi}
        A_B\biggl( pv/R(z)\biggr)
\end{equation}
where $A_B(pv/R(z))$ describes the reduction of the bent-crystal
acceptance with  depth $z$.
The factor $\pi /4$ is exact in a harmonic case;
for the realistic potential one should replace
$\pi$/4 with a slightly different factor $\approx$0.8.

The number of particles dechanneled over the length
$\delta z$, from $z$ to $z+dz$, is equal to
\begin{equation}	\label{b2}
\frac{df(z)}{dz}
 = \frac{2x_c}{d_p}
    \frac{\pi}{4}
    \frac{\theta_{c}}{\Phi}
      \frac{\partial  A_B(pv/R)}{\partial (pv/R)}
      \frac{pv}{R^2(z)} \cdot \frac{dR(z)}{dz}
\end{equation}
All the dechanneled particles are distributed in the angular range
from 0 to the bending angle of the crystal.
The particles dechanneled over $dz$ are exiting in the angular range
$d\theta =dz/R(z)$. Therefore the angular distribution
downstream of the crystal is
\begin{equation}	\label{b3}
\frac{df}{d\theta}
 = R(z)\frac{df(z)}{dz}
 = \frac{2x_c}{d_p}
    \frac{\pi}{4}
    \frac{\theta_{c}}{\Phi}
      \frac{\partial  A_B(pv/R)}{\partial (pv/R)}
      \frac{pv}{R(z)} \cdot \frac{dR(z)}{dz}
\end{equation}
where $z$ and $\theta$ are related through the expression
$d\theta =dz/R(z)$.
For an abrupt ($R'$=$\infty$) or very fast variation
of curvature, Eq. (\ref{b3}) overestimates the phase density,
of course, as we don't take into account a small extra spreading,
$\pm\theta_c$, of the dechanneled particles.
The microscopic
phase density of the particles is to be conserved;
$df/d\theta$ cannot exceed the initial value.

Let us now consider the same beam
incident on the same crystal in the {\em reverse} direction.
That is, the crystal curvature is gradually {\em decreasing}
along the beam path.
Now the particles with the upstream parameters ($x_i, \theta_i$)
equal to the downstream parameters ($x_f, \theta_f$)
of the dechanneled particles from the preceding case,
are captured along the same (reversed) trajectories.
By consideration, the number of particles
which have experienced  transitions {\em from}  the channeled states
in the former case,
is equal to the number of transitions {\em to} the channeled states
in the latter case (as the trajectories are the same).

Therefore the number of particles captured from the interval $d\theta$
is given by Eq.\ (\ref{b3}).
Normalizing it to the number of particles incident on the crystal
in this angular range, $d\theta /2\Phi$,
we obtain
the capture probability:
\begin{equation}	\label{wb2}
w_B(z) = 2\Phi\frac{df(z)}{d\theta}
 = \frac{\pi x_c}{d_p}
    \theta_{c}
      \frac{\partial  A_B(pv/R)}{\partial (pv/R)}
      \frac{pv}{R} \cdot \frac{dR}{dz}
\end{equation}
The above quantity is a function of $z$,
through $R(z)$ and $R'(z)$.
The derivation used just the general arguments,
relating the capture probability to the well-known
function $A_B(pv/R)$ of bending dechanneling.
We cannot show the equivalence of (\ref{wb2}) and (\ref{wb})
for an arbitrary $U(x)$.
For a harmonic $U(x)$, $A_B=(1-R_c/R)^2$ \cite{ufn}.
Then (\ref{wb2}) transforms into
\begin{equation}	\label{wbh2}
w_B  = \frac{2\pi x_c \theta_{c}}{d_p}
\left( 1-\frac{R_c}{R}\right)
\frac{R_cR'}{R}
  = \frac{x_c}{d_p} \lambda
 \left( 1-\frac{R_c}{R}\right)
 \frac{R'}{R}
\end{equation}
which exactly equals the result of previous section, Eq.\ ({\ref{wbh});
the following relations were used here:
 $\lambda$=$2\pi x_c/\theta_{c}$, $R_{c}=pvx_c/2E_c=x_c/\theta_c^2$.
The consistency found between the above two
very different derivations,
Sects. \ref{der1} and \ref{der2},
makes us more confident in the theoretical
predictions.

It may be interesting to say that
the same reversibility approach has been applied
to the {\em scattering} mechanism, leading to a
derivation of the corresponding feed-in probability:
$w_S=\pi R\theta_c/2L_D$ ($L_D$ is the 1/$e$ dechanneling length
due to scattering) \cite{bir-feed1}.

\section{Integral trapping efficiency}

Crystal may trap particles, with probability (\ref{wb},\ref{wb2}),
in the angular range
\begin{equation}	\label{deth}
\Delta\theta = \int^{z_2}_{z_1}dz/R(z) \ ,
\end{equation}
where $z_1$ and $z_2$ are defined by the interval
with $R'>0$ and $R>R_c$.
The capture angle $\Delta\theta$ may be arbitrarily large
(with respective loss in efficiency).
For maximal efficiency, $\Delta\theta$ should be
within the divergence of the incident beam.
From the preceding consideration it is already clear that
the number of particles fed-in from a divergent beam
through the considered mechanism
equals, in the optimal case, to the number trapped
from the same beam at the entry face.
It is interesting, however, to
show explicitly for the harmonic potential, from Eq.\ (\ref{wbh})
that the integral efficiency for a divergent beam coincides with the
entry-face capture efficiency for the same beam,
{\em independent} of the behaviour of function $R(z)$.

The total number of particles, captured
into the channeling mode from the beam
with an angular distribution $P( \theta )$, equals to:
\begin{equation}
N_B =  \int w_B P(\theta )d\theta ,
\end{equation}
where one can integrate over  $z$: $d \theta = dz/R(z)$.
For the uniform distribution $P( \theta ) = 1/2 \Phi$,
{\em covering} the angular
region of capture $\Delta\theta <2\Phi$, we have
\begin{equation}      \label{nb}
N_B =  \int\limits_{z_1}^{z_2} \frac{1}{2\Phi}
           \frac{dz}{R}
           \frac{R' \lambda x_c}{Rd_p}
	   \left( 1-\frac{R_c}{R}\right) =
           \frac{\lambda x_c}{2\Phi d_p}
           \int\limits_{R_1}^{R_2}
           \frac{dR}{R^2}
	   \left( 1-\frac{R_c}{R}\right) ,
\end{equation}
which is independent of the $R(z)$ form.
For the potential well deformation
from   $R_{1} = R_c$   to
   $R_{2} = R$, the number of captured particles equals
\begin{equation}
N_B =   \frac{\lambda x_c}{4\Phi R_{c}d_p}
	   \left( 1-\frac{R_c}{R}\right)^2 =
           \frac{ \pi }{4} \frac{\theta_{c}}{\Phi}
	   \frac{2x_c}{d_p}
	   \left( 1-\frac{R_c}{R}\right)^2  ,
\end{equation}
which is exactly equal to the efficiency
for the entry-face capture.

This conclusion, derived analytically in the harmonic approximation,
is true also in the general case  of an arbitrary potential,
as a consequence of the
Liouville's theorem. In the absense of dissipative processes
the microscopic
phase density of the captured beam equals the phase density of the
incident beam.

The angular acceptance of a crystal deflector
could be made much wider than the divergence of an incident beam,
$\Delta\theta\gg 2\Phi$.
In this case the intensity decreases according to (\ref{nb}), where
$R_1$ and $R_2$ are the bending radii at the points,
where the beam envelopes are
tangent to the crystal planes.

\section{Experimental evidence}

The first experimental indication of the role of the gradient effect in
volume capture was made at 70 GeV \cite{ches90}.
The measurements gave for $w_B$ the values factor of 2 higher
than those predicted by Eq.\ (\ref{wb}).
The dependence of $w_B$ on $R$ was measured to be that
predicted by Eq.\ (\ref{wb})
(linear in case of Ref.\ \cite{ches90}).

\section{Additivity of the two mechanisms of feed-in}

In bent crystals both mechanisms, bending and scattering,
may contribute to the feeding-in.
In the following we present a qualitative analysis, which
indicates that the two mechanisms contribute in an additive way.

The 'scattering feeding' is efficient only at the tangential points.
The particles pass over the potential well
with the initial transverse energies being in the range
from $U_0$ to $U_0+\Delta U$.
Having scattered during this oscillation,
the particles get some distribution $f(E_T)$
with a (interesting to us) tail to smaller $E_T<U_0$.
Those particles which have $E_T<E_c$, are trapped in
the stable channeled states.
Their number is
\begin{equation}
N_S =  \int_0^{E_c} f(E_T)dE_T ,
\end{equation}
The curvature variability modifies $E_c$; then the trapped fraction is
\begin{equation}
N =  \int_0^{E_c+\delta E_c} f(E_T)dE_T =
     N_S + \int_{E_c}^{E_c+\delta E_c} f(E_T)dE_T
\end{equation}
Since normally $\delta E_c\ll E_c$ over one oscillation $\lambda$,
we have
\begin{equation}
N =   N_S + f(E_c)\delta E_c = N_S + N_B .
\end{equation}
The term $N_B=f(E_c)\delta E_c\approx \delta E_c/\Delta U$
is the fraction of particles trapped through the bending mechanism,
as it is defined in Section \ref{der1}.

For positive  $\delta E_c$ (the well grows),
the two contributions are summed: $N=N_S+N_B$.
For negative  $\delta E_c$ (the well shrinks),
the contributions are subtracted: $N=N_S-N_B$.

\section{Conclusion}

Although the processes considered cannot increase
the beam deflection efficiency,
they contribute to a wide angular acceptance ($\gg\theta_c$)
crystal deflector. The deflection efficiency (always smaller
than in the regular case of the entry-face capture) of such deflector
can be {\em designed}, and be varied in a broad range of values.
Moreover, this efficiency is independent of the incident beam divergence
(in contrast to the regular case).
This may be valuable, for instance, for the proposed \cite{doble}
measurements of CP-violation with use of $K_S$ and $K_L$ meson beams
at CERN, where beam attenuation, stability, and low background are
important issues.

Obviously, the secondary particles produced in collisions with crystal
nuclei can be captured into the channeling mode, from the crystal bulk,
through the feed-in processes only.
This may be applicable, e.g., in some ideas of crystal usage
for the experiments in particle physics \cite{relchan}.

As the energy increases, the rate of transitions
from the random to the channeled states changes.
For a given crystal,
the rate $w_S$ of the scattering-induced transitions
decreases as 1/$(pv)^{3/2}$
since the scattering vanishes \cite{bir-feed1}.
In contrast, the rate $w_B$ of bending capture,
Eq.\ (\ref{wb}), grows as $(pv)^{1/2}$
since the centrifugal effects strengthen.
The ratio of the two feeding mechanisms
changes in favour of bending with the energy increase.

At sufficiently high energies (above $\sim$1 TeV),
the processes of feeding-out and feeding-in in bent crystals
are predominantly defined by the centrifugal effects.
These processes are described by the continuum model,
which is a well established analytical theory.
Furthermore, these centrifugal processes are easily controlled,
through the proper design of the $R(z)$-function.

\begin{figure}
\begin{center}
\setlength{\unitlength}{2mm}
\begin{picture}(25,50)(-15,-20)
\thicklines

\put(-21.0,-1.08){\circle{.5}} \put(-21.0,  9.419){\circle*{.5}}
\put(-20.0, 2.26){\circle{.5}} \put(-20.0,  12.24){\circle*{.5}}
\put(-18.9, 3.80){\circle{.5}} \put(-18.9,  13.25){\circle*{.5}}
\put(-17.8, 3.27){\circle{.5}} \put(-17.8,  12.20){\circle*{.5}}
\put(-16.8, 1.47){\circle{.5}} \put(-16.8,  9.876){\circle*{.5}}
\put(-15.8,-0.53){\circle{.5}} \put(-15.8,  7.342){\circle*{.5}}
\put(-14.7,-2.19){\circle{.5}} \put(-14.7,  5.157){\circle*{.5}}
\put(-13.7,-3.42){\circle{.5}} \put(-13.7,  3.405){\circle*{.5}}
\put(-12.6,-4.27){\circle{.5}} \put(-12.6,  2.026){\circle*{.5}}
\put(-11.6,-4.82){\circle{.5}} \put(-11.6,  0.951){\circle*{.5}}
\put(-10.5,-5.12){\circle{.5}} \put(-10.5,  0.124){\circle*{.5}}
\put(-9.40,-5.22){\circle{.5}} \put(-9.40, -0.494){\circle*{.5}}
\put(-8.40,-5.14){\circle{.5}} \put(-8.40, -0.938){\circle*{.5}}
\put(-7.40,-4.90){\circle{.5}} \put(-7.40, -1.228){\circle*{.5}}
\put(-6.30,-4.53){\circle{.5}} \put(-6.30, -1.383){\circle*{.5}}
\put(-5.20,-4.04){\circle{.5}} \put(-5.20, -1.415){\circle*{.5}}
\put(-4.20,-3.43){\circle{.5}} \put(-4.20, -1.334){\circle*{.5}}
\put(-3.10,-2.72){\circle{.5}} \put(-3.10, -1.148){\circle*{.5}}
\put(-2.10,-1.91){\circle{.5}} \put(-2.10, -0.861){\circle*{.5}}
\put(-1.05,-1.00){\circle{.5}} \put(-1.05, -0.478){\circle*{.5}}
\put( 1.05, 1.09){\circle{.5}} \put( 1.05,  0.572){\circle*{.5}}
\put( 2.10, 2.29){\circle{.5}} \put( 2.10,  1.239){\circle*{.5}}
\put( 3.10, 3.57){\circle{.5}} \put( 3.10,  2.003){\circle*{.5}}
\put( 4.20, 4.96){\circle{.5}} \put( 4.20,  2.868){\circle*{.5}}
\put( 5.20, 6.46){\circle{.5}} \put( 5.20,  3.838){\circle*{.5}}
\put( 6.30, 8.07){\circle{.5}} \put( 6.30,  4.920){\circle*{.5}}
\put( 7.40, 9.80){\circle{.5}} \put( 7.40,  6.126){\circle*{.5}}
\put( 8.40, 11.6){\circle{.5}} \put( 8.40,  7.467){\circle*{.5}}
\put( 9.40, 13.6){\circle{.5}} \put( 9.40,  8.961){\circle*{.5}}
\put( 10.5, 15.8){\circle{.5}} \put( 10.5,  10.63){\circle*{.5}}
\put( 11.6, 18.2){\circle{.5}} \put( 11.6,  12.51){\circle*{.5}}
\put( 12.6, 20.9){\circle{.5}} \put( 12.6,  14.63){\circle*{.5}}
\put( 13.7, 23.8){\circle{.5}} \put( 13.7,  17.06){\circle*{.5}}
\put( 14.7, 27.2){\circle{.5}} \put( 14.7,  19.86){\circle*{.5}}
\put( 15.8, 30.9){\circle{.5}} \put( 15.8,  23.10){\circle*{.5}}
\put( 16.8, 35.0){\circle{.5}} \put( 16.8,  26.68){\circle*{.5}}
\put( 17.8, 39.0){\circle{.5}} \put( 17.8,  30.07){\circle*{.5}}
\put( 18.9, 41.6){\circle{.5}} \put( 18.9,  32.17){\circle*{.5}}
\put( 20.0, 42.1){\circle{.5}} \put( 20.0,  32.21){\circle*{.5}}
\put( 21.0, 40.9){\circle{.5}} \put( 21.0,  30.43){\circle*{.5}}

\put(-24,36) {\vector(1,0){46}}
\put(-24,22) {\vector(1,0){36}}
\put(12,23) {\vector(-1,0){36}}
\put(-24,8) {\vector(1,0){29}}
\put(5,9) {\vector(-1,0){20}}

\put(-28,35) {\large 3)}
\put(-28,21) {\large 2)}
\put(-28,7)  {\large 1)}

\put(-35,-10) {\vector(1,0){55}}
\put(-35,-10) {\vector(0,1){55}}
\put(-39,29){\Large U}
\put(-33,41){\Large $U_0+\Delta U$}
\put(-33,2){\Large $U_0$}
\put(15,-8){\Large x}
\put(-36,4) {\line(1,0){2}}
\put(-36,42) {\line(1,0){2}}

\end{picture}
\end{center}
\caption{
  Schematic picture of bending capture.
  The effective interplanar potential $U_{eff}(x)$
  is shown at some coordinate $z$ (o),
  and at $z+\lambda$ ($\bullet$).
  There are
  three kinds of particles: (1) captured, (2) reflected,
  (3) transmitted and interacted with the next wall.
}
  \label{gap}
\end{figure}
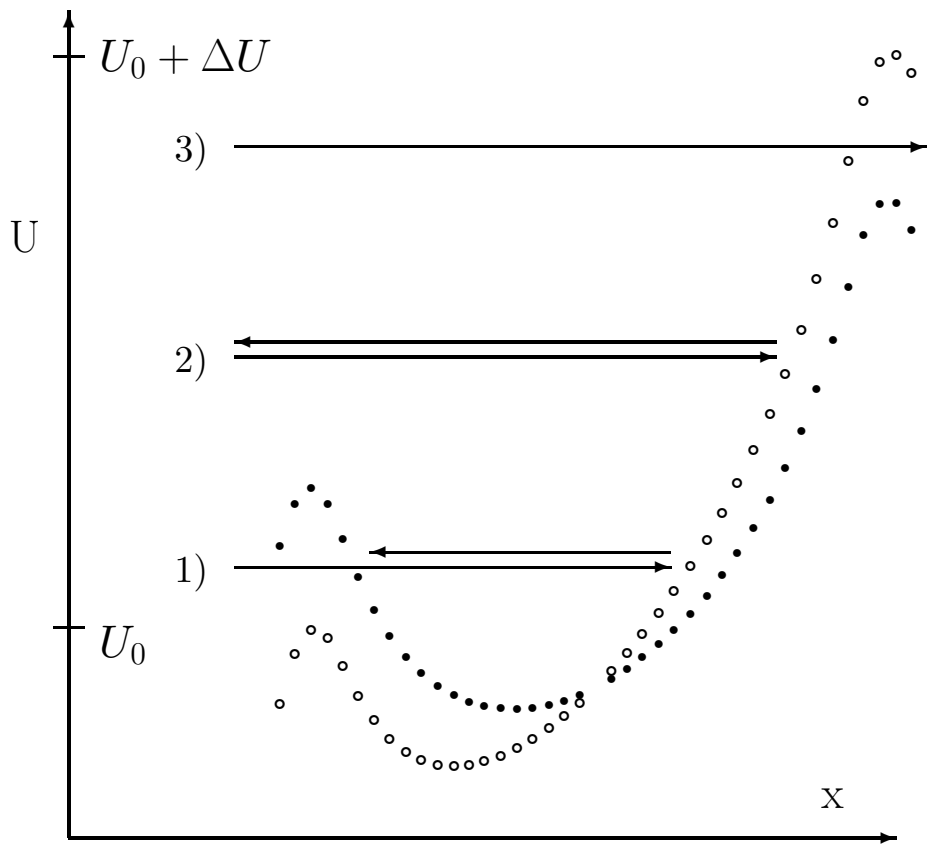

\end{document}